\documentclass[%
amsmath,amssymb,
aps,superscriptaddress,prl,notitlepage,reprint
]{revtex4-1}

\usepackage{bm}
\usepackage{amsmath, amsfonts,amssymb,mathrsfs}
\usepackage{physics}
\usepackage{color}
\usepackage{soul}
\usepackage{graphicx}

\providecommand{\bp}{{\bm{p}}}

\providecommand{\bk}{{\bm{k}}}
\providecommand{\bq}{{\bm{q}}}

\begin{document}
\title{Bose Condensation of Photons Thermalized via Laser Cooling of Atoms}

\author{Chiao-Hsuan Wang}
\affiliation{Joint Center for Quantum Information and Computer Science, NIST/University of Maryland, College Park, Maryland 20742, USA}
\affiliation{Joint Quantum Institute, NIST/University of Maryland, College Park, Maryland 20742, USA}

\author{M. J. Gullans}
\affiliation{Department of Physics, Princeton University, Princeton, New Jersey 08544, USA}

\author{J. V. Porto}
\affiliation{Joint Quantum Institute, NIST/University of Maryland, College Park, Maryland 20742, USA}

\author{William D. Phillips}
\affiliation{Joint Quantum Institute, NIST/University of Maryland, College Park, Maryland 20742, USA}

\author{Jacob M. Taylor}
\affiliation{Joint Center for Quantum Information and Computer Science, NIST/University of Maryland, College Park, Maryland 20742, USA}
\affiliation{Joint Quantum Institute, NIST/University of Maryland, College Park, Maryland 20742, USA}

\begin{abstract}
A Bose-Einstein condensate (BEC) is a quantum phase of matter achieved at low temperatures.  Photons, one of the most prominent species of bosons, do not typically condense due to the lack of a particle number-conservation.  We recently described a photon thermalization mechanism which gives rise to a grand canonical ensemble of light with effective photon number conservation between a subsystem and a particle reservoir.  This mechanism occurs during Doppler laser cooling of atoms where the atoms serve as a temperature reservoir while the cooling laser photons serve as a particle reservoir. 
Here we address the question of the possibility of a BEC of photons in this laser cooling photon thermalization scenario and theoretically demonstrate that a Bose condensation of photons can be realized by cooling an ensemble of two-level atoms (realizable with alkaline earth atoms) inside a Fabry-Perot cavity. 
\end{abstract}
\maketitle

\section{Introduction}
A Bose-Einstein condensation (BEC) is a striking example of quantum behavior where a macroscopic number of bosons occupy the same single-particle state.
Traditionally, BEC occurs in systems with particle number conservation, either represented by a grand canonical ensemble (GCE) or in a system closed to particle exchange. Thus, we would expect photons, whose number is not conserved and which do not generally admit a GCE description, not to condense. For example, when one cools a blackbody, photons disappear; instead of forming a condensate, one reaches a vacuum state at $T=0$.

There are several exceptions to this, however.  For example, light can acquire nonzero chemical potential and form a BEC via mutual interactions mediated by matter in the form of hybridized light-matter particles called polaritons~\cite{Imamog1996,Deng2002,Kasprzak2006,Balili2007,Keeling2007,Deng2010,Fleischhauer2008}, photons in a plasma~\cite{ZelDovich1969,Mendonca2017}, cavity photons in a nonlinear resonator~\cite{Chiao2000}, and propagation of light in a nonlinear medium~\cite{Connaughton2005,Sun2012,Chiocchetta2016,Santic2018}. Photons can also thermalize with a number-conserving reservoir, and condense~\cite{Snoke2013}, in a dye-filled microcavity~\cite{Klaers2010a,Klaers2010b,Klaers2012,DeLeeuw2013,Kirton2013,Schmitt2014,Marelic2015,Greveling2017,Walker2017}, an optomechanical cavity~\cite{Weitz2013,Fani2016}, an ideal gases composed of two kinds of atoms~\cite{Kruchkov2013}, a 1D microtube~\cite{Kruchkov2016}, and a fiber~\cite{Weill2017}. In all of these cases, the average photon number is approximately conserved either by photon confinement in a cavity or through the compensation of loss via nonequilibrium pumping.

On the other hand, interactions between atoms and optical cavities have made possible novel atom-cooling mechanisms~\cite{Heinzen1987,Vuletic2000,Vuletic2001a,Curtis2001,Beige2005,Lev2008,Seletskiy2010,Xu2016,Hosseini2017} as well as peculiar states of light~\cite{Baumann2010}.  We recently found a different photon thermalization mechanism that occurs in Doppler laser cooling of a high optical depth atomic ensemble~\cite{Wang2018a}, which requires neither matter-matter nor effective photon-photon interactions.  Here we show that this thermalization mechanism can lead to Bose condensation of photons.  Specifically, in our scenario the laser-cooled atoms serve as a thermal reservoir while the laser photons serve as a particle reservoir for the reemitted photons, leading to a grand canonical ensemble of photons at the atomic temperature and with a chemical potential very close to the energy of a single laser photon.

To give a practical setting for our work, we adopt the now standard approach to controlling the photon dispersion relation by using a Fabry-Perot cavity where transverse excitations of a single longitudinal mode can be mapped onto a 2D massive bosonic gas with a harmonic trapping potential.  While previous theoretical analysis of BEC has been mostly focused on the identification of a critical temperature or critical number (density)~\cite{Bagnato1991,Mullin1997,Klaers2010a,Kirton2013,DeLeeuw2013,Klaers2014}, here we consider the photon condensate fraction as a function of temperature and chemical potential (set effectively by the cooling laser detuning from the cavity).  By carefully treating the modification due to loss, we are able to construct a phase diagram as a function of laser frequency and field strength, showing condensate, thermal, quasithermal and gain regimes for cavity photons with calculated values appropriate for the Yb intercombination transition.

\section{Photon Thermalization}

\begin{figure}[htbp]
\begin{center}
\includegraphics[width=\linewidth]{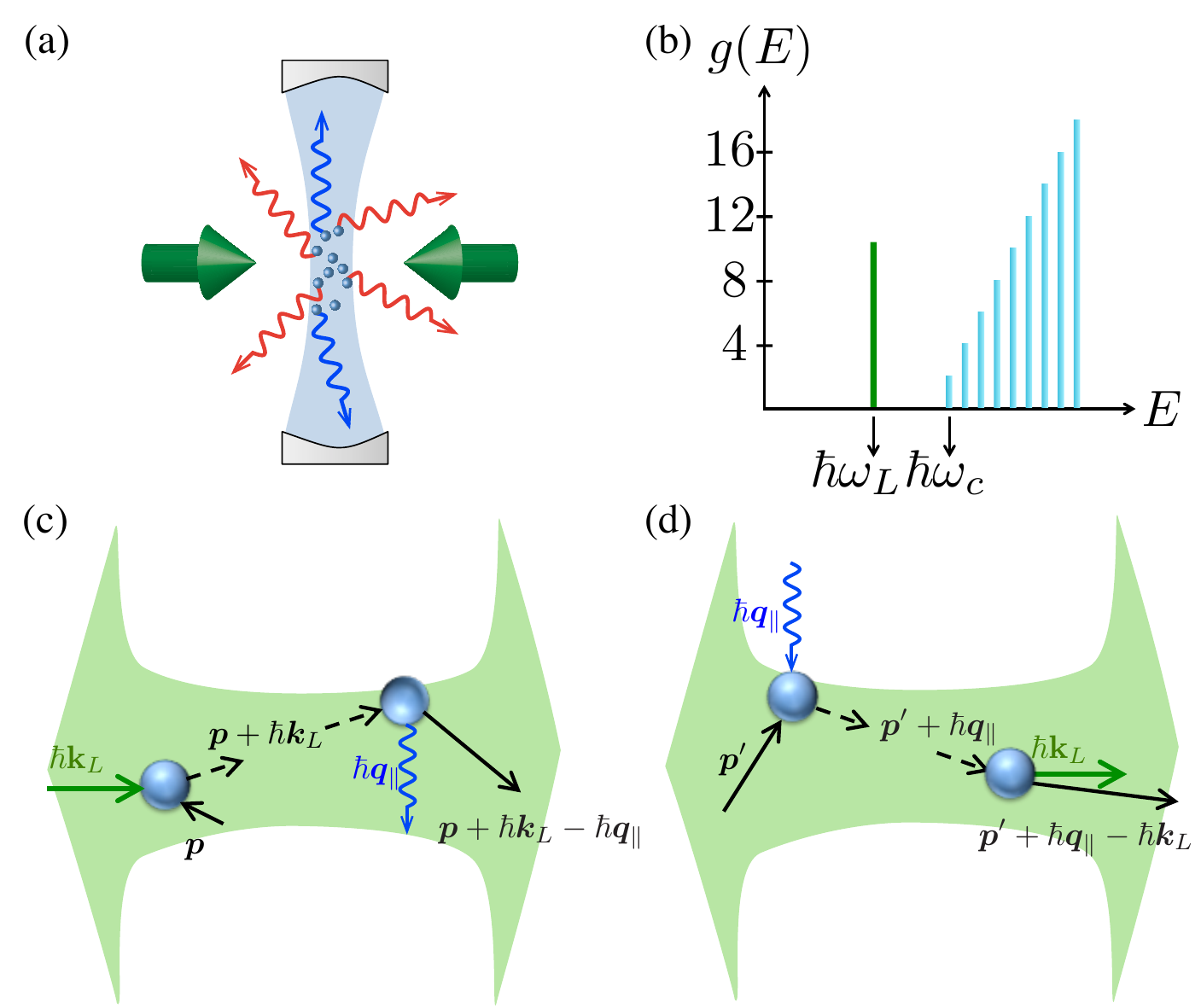}
\caption{(a) Schematic of an ensemble of two-level atoms which are Doppler-cooled by laser fields (green arrows) and free-space photon modes (red arrows), while also interacting with cavity photon modes (blue arrows within the light-blue region).  
(b) The state degeneracy $g(E)$ (equivalent to density of states) of the transverse cavity modes is equivalent to that of a 2D massive particle in a harmonic trapping potential  with a lower-cutoff energy $\hbar \omega_c=\hbar\omega_{q_{\parallel} 00}$ including polarization. The green line shows the energy of a single laser photon $\hbar \omega_L$. 
(c) The scattering process in which an atom is excited by the laser field and then emits a cavity photon. (d) The scattering process in which an atom absorbs a cavity photon and then scatters back into the laser field.
}
\label{fig:cavity} 
\end{center}
\end{figure}

Consider 3D Doppler cooling of noninteracting two-level atoms in a long cavity [i.e. the cavity subtends a small solid angle as illustrated in Fig.~\ref{fig:cavity}(a)]. The cavity separates the emitted photons into long-lived cavity modes and lossy, free-space modes. When the atom is excited by a laser photon, it is most likely to de-excite by emitting a photon into free-space, and this scattering process induces Doppler cooling of atoms~\cite{Chu1998,Cohen-Tannoudji1998,Phillips1998}. A rarer event is the spontaneous emission into the cavity. However, the high quality of the cavity mirrors allows those cavity photons to be reabsorbed by the atoms and preferentially emitted into the cooling beam, if the cooling laser is sufficiently intense.  Both processes produce light whose coherence is described as thermal, in the quantum optics sense of having a photon autocorrelation that is peaked at short times. However, multiple scattering also leads to photon thermalization in an energetic sense across different transverse cavity modes.  Photons thermalize with the atomic motion in an approximate particle number-conserving way where the cooling laser acts as a photon reservoir.

In the low excitation limit for Doppler cooling, $\Omega^2 \ll \abs{\bar{\Delta}_L+ i\Gamma/2}^2$~\cite{Cohen-Tannoudji1992}, the scattering between the red-detuned laser fields and the free-space modes will lead to cooling of atoms to a temperature $k_B T = \beta^{-1} =\hbar(\bar{\Delta}_L^2+\Gamma^2/4)/2 |\bar{\Delta}_L|$ \cite{Lett1989}. Here $2 \Omega$ is the Rabi frequency of the cooling laser field, $\bar{\Delta}_L= \omega_L-\omega_A-\frac{\hbar\bk_L^2}{2m_A} <0$ is the laser detuning including the recoil shift $\frac{\hbar\bk_L^2}{2m_A}$; $\Gamma$, $\omega_A$, and $m_A$ are the natural linewidth, two-level transition frequency, and the mass of the atoms; $\omega_L$ and $\hbar\bk_L$ are the frequency and momentum of the laser photons.

In addition to the timescale set by photon scattering rate in the Doppler-cooling process there is the slower dynamics associated with emission and absorption of the cavity photons by atoms. In the large detuning and high power limit $\abs{\bar{\Delta}_L}^2 \gg \Omega^2 \gg \Gamma^2$, the dominant processes involving creation and annihilation of cavity photons are the scatterings between a laser photon and a cavity photon as illustrated in Fig.~\ref{fig:cavity}(c)-(d).
For a single longitudinal mode with a longitudinal momentum $\bq_\parallel = q_\parallel \hat{z} $, the transverse modes of the cavity can be expressed in terms of Laguerre-Gauss modes labeled by the radial index $l \in \mathbb{N}$ and azimuthal index $m \in \mathbb{Z}$~\cite{Kogelnik1966,Siegman1986}. According to Fermi's golden rule, the total rate that an atom scatters laser photons into a cavity mode with mode frequency $\omega_{q_{\parallel}lm}$~\cite{Kogelnik1966} [Fig.~\ref{fig:cavity}(c)] is, in the large detuing limit, 
\begin{align}
(n_{q_{\parallel}lm} +1 )\Lambda^+_{q_{\parallel}lm,L} \approx& \sum_\bp \frac{2 \pi}{\hbar} \frac{\hbar^2 \Omega^2 \alpha_{q_\parallel}^2}{\abs{\bar{\Delta}_L}^2} (n_{q_{\parallel}lm} +1 )\notag\\ &\times
\delta \left(\hbar \omega_L + K(\bp)-\hbar\omega_{q_{\parallel}lm} - K(\bp{'}) \right),
\label{Lambda+L}
\end{align}
where $\alpha_{q_\parallel}^2$ is the spatial average of $\alpha_{q_\parallel lm}^2$ ($2\alpha_{q_\parallel l m}$ is the single-photon Rabi frequency of the transverse cavity mode $q_\parallel lm$),  $n_{q_{\parallel}lm}$ is the cavity photon occupation number, $\Lambda_{q_{\parallel}lm,L}^+$ is the single-cavity-photon emission rate mediated by the laser, $\bp$ and $\bp{'}=\bp+\hbar \bk_L-\hbar \bq_{\parallel}$ are the atomic momentum before and after the scattering event, and $K(\bp)=\bp^2/2m_A$ is the kinetic energy of the atom.  We are working in the paraxial limit so that $q_{q_{\parallel}lm} \equiv \omega_{q_{\parallel}lm}/c \approx q_{\parallel}$.  Furthermore, the atoms are taken to have a uniform spatial distribution within the cavity mode volume so that the spatial average of $\alpha^2_{q_{\parallel}lm}$ is independent of $l$ and $m$.

Similarly, the total rate that an atom scatters cavity photons into the laser field [Fig.~\ref{fig:cavity}(d)] is
\begin{align}
 n_{q_{\parallel}lm}\Lambda^-_{q_{\parallel}lm,L} \approx& \sum_{\bp{'}} \frac{2 \pi}{\hbar} \frac{\hbar^2 \Omega^2 \alpha_{q_\parallel} ^2}{\abs{\bar{\Delta}_L}^2} n_{q_{\parallel}lm}\notag\\ &\times\delta \left(\hbar \omega_{q_{\parallel}lm} + K(\bp{'})-\hbar\omega_L - K(\bp) \right),
\label{Lambda-L}
\end{align}
where $\Lambda_{q_{\parallel}lm,L}^-$ is the single-cavity-photon absorption rate mediated by the laser, $\bp{'}$ is the atomic momentum before the scattering event, and $\bp=\bp{'}+\hbar \bq_{\parallel} -\hbar \bk_L$ is the atomic momentum after the scattering event.

Equilibration between emission and absorption of cavity photons mediated by the cooling laser will lead to a detailed balance condition such that Eq.~(\ref{Lambda+L}) equals Eq.~(\ref{Lambda-L}), which gives
\begin{align}
\frac{\bar{n}_{q_{\parallel}lm}+1}{\bar{n}_{q_{\parallel}lm}}=\frac{\Lambda^-_{q_{\parallel}lm,L}}{\Lambda^+_{q_{\parallel}lm,L}}
= \sum_i \frac{e^{-\beta K(\bp_i{'})}}{e^{-\beta K(\bp_i)}} =e^{\beta \hbar(\omega_{q_{\parallel}lm} -\omega_L)}.
\label{GrandL}
\end{align}
Here $\bar{n}_{q_{\parallel}lm}$ is the mean number of photons under detailed balance. The Boltzman factor is picked up by each pair of $\bp_i$ and $\bp_i{'}$ satisfying the energy conservation condition $K(\bp_i{'})-K(\bp_i)=\hbar(\omega_L-\omega_{q_{\parallel}lm})$ when summing over the atomic momentum distribution.
This equilibration condition can be understood within the framework of photon thermalization with a parametrically coupled bath \cite{Hafezi2015,Gullans2016,Wang2018,Wang2018a}, where the conservation of the total number of cavity plus laser photons during the scattering processes imposes a nonzero chemical potential $\hbar \omega_L$ to the cavity photons. 
For $\omega_{q_{\parallel}lm} > \omega_L$, one has $\bar{n}_{q_{\parallel}lm}=\frac{1}{e^{\beta \hbar(\omega_{q_{\parallel}lm} -\omega_L )}-1}$ corresponding to a grand canonical distribution; for $\omega_{q_{\parallel}lm} < \omega_L$, one expects gain or lasing instead of an equilibrium steady state since $\Lambda^{+}_{q_{\parallel}lm,L} > \Lambda^{-}_{q_{\parallel}lm,L}$.

In reality, cavity photons also suffer from losses either due to scattering into the free space modes or dissipations at the cavity mirrors.  Based on the theoretical tools developed in Ref.~\cite{Wang2018a}, assuming perfect cavity mirrors, the detailed balance condition is modified by the loss caused by scattering of the cavity photons into the free-space modes to
\begin{align}
&\frac{\bar{n}_{q_{\parallel}lm}+1}{\bar{n}_{q_{\parallel}lm}}
\approx \,\, e^{\beta\hbar(\omega_{q_{\parallel}lm}-\omega_L)}\notag\\&+\frac{\Gamma}{\Omega^2}\frac{|\bk_L-\bq_{\parallel}|}{\sqrt{2 \pi \beta m_A}}e^{\frac{\beta}{2m_A}\left(-\frac{m_A}{|\bk_L-\bq_{\parallel}|}(\omega_{q_{\parallel}lm}-\omega_L)-\frac{\hbar|\bk_L-\bq_{\parallel}|}{2}\right)^2}.
\label{DetailedBalance}
\end{align}

This loss-modified result represents a small correction to the grand-canonical form Eq.~(\ref{GrandL}) in the high power limit $\Omega \gg \Gamma$, which is the focus of this work. Furthermore, the correction depends on the cavity mode frequency,  and is larger for larger transverse cavity modes (when the frequency  difference between the given cavity mode and the laser,  $\omega_{q_{\parallel lm}}-\omega_L$, is larger).  At large but finite power, we can incorporate the corrections from the second term on the right hand side of Eq.~(\ref{DetailedBalance}) into a shifted chemical potential $\hbar \omega_L-\delta \mu$ and a mode-dependent effective temperature $\beta_{\rm eff}^{-1}$, where, formally, $\delta \mu$ and $\beta_{\rm eff}$ are  defined by the equations~\cite{Wang2018a}
\begin{align}
 1 = e^{-\beta \delta \mu}+\frac{\Gamma}{\Omega^2}\frac{|\bk_L-\bq_{\parallel}|}{\sqrt{2 \pi \beta m_A}}e^{\frac{\beta }{2m}\left(\frac{m_A \delta \mu}{\hbar|\bk_L-\bq_{\parallel}|}-\frac{\hbar|\bk_L-\bq_{\parallel}|}{2}\right)^2},
 \label{deltamu}
 \end{align}
\begin{align}
 e^{\beta_{\rm eff}(\hbar\omega_{q_{\parallel}lm}-\hbar\omega_L+\delta \mu)}= \frac{\bar{n}_{q_{\parallel}lm}+1}{\bar{n}_{q_{\parallel}lm}}.
\label{Teff}
\end{align}
Equation (\ref{Teff})  predicts a  transition from equilibrium to gain at the shifted frequency $\omega_{q_{\parallel}lm}=\omega_L-\delta \mu /\hbar$. For low power, $\Omega < \Omega_c$ for some critical value $\Omega_c$, photon loss is large enough such that
Eq.~(\ref{deltamu}) has no solutions.  In that case, $\delta \mu$ and $\beta_{\rm eff}$ are no longer well-defined and only quasithermal light (where the photon distribution cannot be described by a single well-defined temperature) is expected.

\section{2D photon BEC in a cavity}
Restricting the cavity photon states to $\omega_{\bq} \geq \omega_c >\omega_L-\delta \mu/\hbar$, for a cavity cut-off frequency $\omega_c$ equal to the lowest transverse mode frequency, can prevent the regime of gain and take us towards a photon BEC. Specifically, we control the photon density of states with a cavity and, further, consider a Fabrey-Perot cavity with curved mirrors to realize a quadratic dispersion relation for the energy of photons, as has been used to create a BEC of light~\cite{Chiao2000,Klaers2011}.  In contrast to prior work, here we consider a long cavity subtending a small solid angle, which makes the atoms emit mostly into the free-space modes, enabling Doppler cooling [Fig.~\ref{fig:cavity}(a)].

Specifically, the frequency of a cavity photon in a Laguerre-Gauss mode $(q_{\parallel},l,m)$ is given by~\cite{Kogelnik1966,Siegman1986}
\begin{align}
    \omega_{q_{\parallel}lm} \equiv c q_{\parallel}+ \frac{c}{D_0} (2l +\abs{m}+1) \cos^{-1}\left(1-\frac{D_0}{R}\right),
    \label{omegaqlm}
\end{align}
where $D_0$ is the distance between cavity mirrors and $R$ is the radius of curvature of the mirrors.  The transverse energy spectrum and the density of states of a single longitudinal mode inside the cavity is identical to that of a Hamiltonian for a (fictitious) massive 2D particle in a harmonic potential trap:
\begin{align}
\hat{H}_{\rm \perp} =\frac{(\hbar \hat{\bq}_{\perp})^2}{2 M_{\rm ph}}+\frac{1}{2} M_{\rm ph} \omega_T^2 \hat{r}_{\perp}^2,
\label{Eph}
\end{align}
where $M_{\rm ph}=\hbar q_{\parallel}/c$ is the mass of the 2D particle, $\hbar \hat{\bq}_{\perp}$ and $\hat{r}_{\perp}$ are the corresponding momentum and position operators, and the trapping frequency is $\omega_T= \frac{c}{D_0} \cos^{-1}(1-D_0/R)$.

As with 2D massive bosons in a harmonic trap, the cavity photons can undergo Bose condensation into the ground mode (the lowest transverse mode, which sets the cut-off frequency $ \omega_c \equiv  \omega_{q_{\parallel}00}$ to the cavity modes). To make a direct connection to the typical BEC theory, we define a displaced chemical potential $\mu=\hbar(\omega_L-\omega_c)$ to compare the original chemical potential with the ground mode energy, such that in the absence of loss the whole system is in thermal equilibrium for $\mu<0$, achieves BEC in the the thermodynamic limit\footnote{The thermodynamic limit is reached by taking $n_{\rm tot}\to \infty$ while keeping the 2D number density inside the trap, which is proportional to $n_{\rm tot} \omega_T^2$, fixed.} at $\mu=0$ , and exhibits gain when $\mu>0$.  The critical temperature of condensation is given by $T_c \approx \hbar \omega_T \sqrt{3 n_{\rm tot}}/\pi k_B$ where $n_{\rm tot}$ is the steady-state average photon number~\cite{Bagnato1991,Mullin1997,Klaers2011}.
In contrast to many prior theoretical discussions of trapped atomic and photon BEC, 
there are two distinguishing features in photon BEC transitions under this laser cooling scenario. First, our system is better treated in the context of a grand canonical ensemble with a controlled chemical potential. Second, the energy-dependent loss mechanisms can affect the transition.  We will first explore lossless BEC physics under the framework of a number reservoir with a controlled chemical potential, and later include the effect of loss.

In our laser cooling scenario, one can control $T$ and $\mu$ independently by setting an approximately fixed temperature $k_B T \approx \hbar \abs{\bar{\Delta}_L}/2$ for a large detuning from the atomic transition, and adjusting $\mu$ by the small laser detuning from $\omega_c$. The analogous 2D massive bosonic gas experiences a fixed trapping frequency $\omega_T$ determined by the geometry of the cavity. We note that $n_{\rm tot}$ is determined jointly by $T$, $\mu$, and $\omega_T$, and we explore the BEC transition in the context of a fixed $T$ and $\omega_T$ while varying $\mu$.

\begin{figure}[htbp]
\begin{center}
\includegraphics[width=\linewidth]{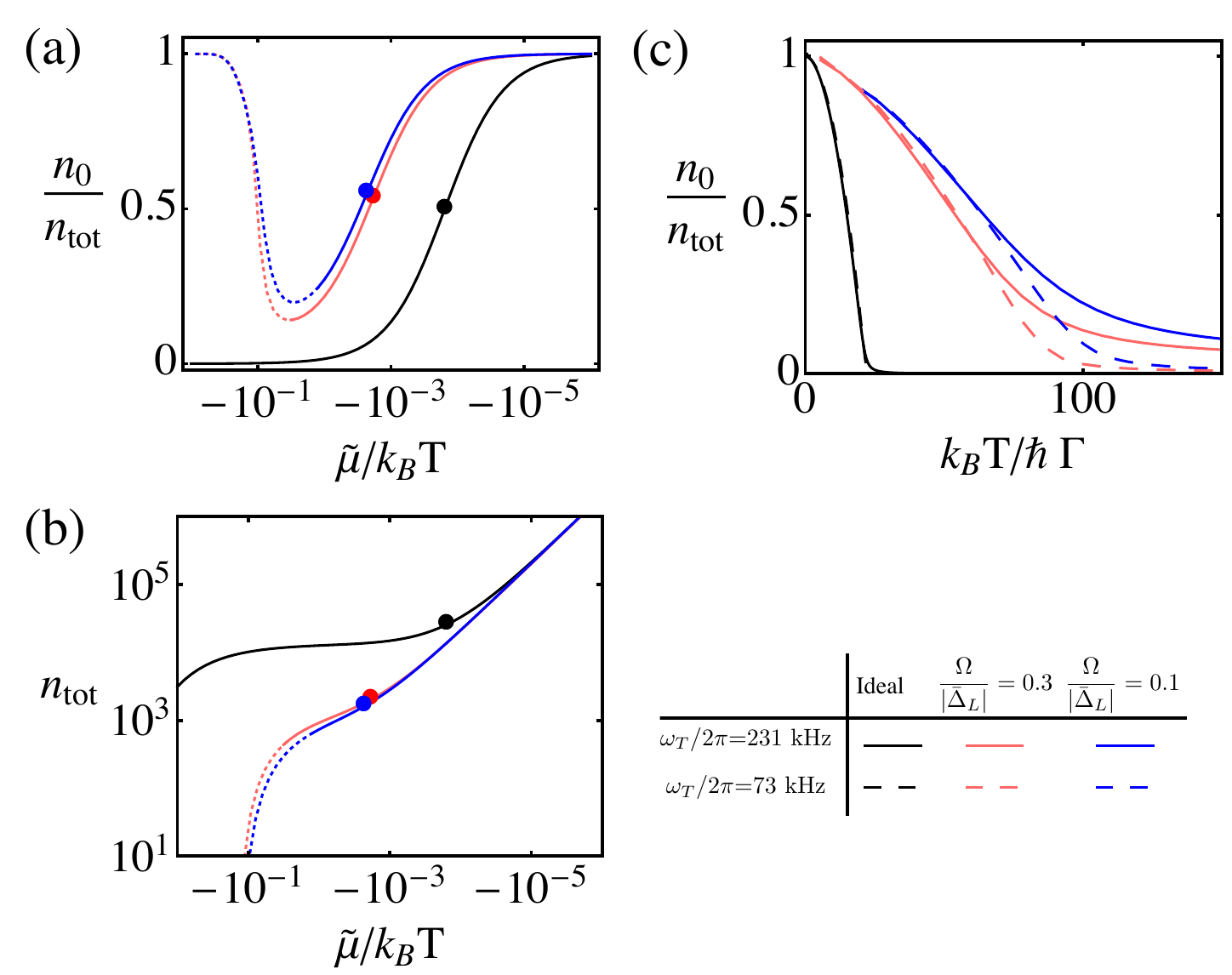}
\caption{(a),(b) The condensate fraction $n_0/n_{\rm tot}$ [shown in (a)] and the  total average number of photons $n_{\rm tot}$ [shown in (b)] at a fixed temperature as a function of the loss-shifted chemical potential $\tilde{\mu}$ plotted on a logarithmic scale. The ideal grand canonical ensemble result is shown in black lines, and the colored lines represent the modified results with two different Rabi frequencies. The critical points representing the onset of the BEC phase are identified by dots. We take parameters for the $~^1 S_0 - ~^3 P_1$ Yb intercombination transition~\cite{Ludlow2015}, $\omega_L/2\pi \approx \omega_A/2\pi =$ 539 THz, $\Gamma/2\pi=$ 180 kHz, $E_{r,\bk_L}/h=$ 3.74 kHz, and assume $|\bk_L-\bq| \approx \sqrt{2} |\bk_L|$, $\bar{\Delta}_L \approx - 157 \Gamma$, and a cavity trapping frequency $\omega_T/2 \pi$= 231 kHz.  (c) The condensate fraction $n_0/n_{\rm tot}$ as a function of temperature while keeping the total number of photons fixed.  The ideal grand canonical distribution is shown in black lines while the colored lines indicate loss-modified results with two different Rabi frequencies. We take parameters for the Yb intercombination transition with  $\omega_T/2\pi=231$ kHz for $n_{\rm tot}=10^3$ and $\omega_T/2 \pi=73$ kHz for $n_{\rm tot}=10^4$ such that $n_{\rm tot}(\omega_T/2\pi)^2=5.33 \times 10^{13} \text{s}^{-2}$, proportional to the 2D number density, is fixed, which leaves the critical temperature essentially unchanged.
}
\label{fig:fractot}
\end{center}
\end{figure}

For the ideal (lossless) grand canonical ensemble of a 2D massive Bose gas in a harmonic trap, $n_{\rm tot}$ is
\begin{align}
n_{\rm tot}=\sum_{l=0}^{\infty}\sum_{m=-\infty}^{\infty} \frac{2}{e^{\beta \left\{ (2l+\abs{m}) \hbar\omega_T - \mu\right\}}-1}=\sum_{j=0}^{\infty} \frac{2(j+1)}{e^{\beta  (j \hbar\omega_T - \mu)}-1},
 	\label{ntot}
\end{align}  
where the factor of 2 comes from polarization degeneracy and $j=2l+\abs{m}$. Each cavity mode with frequency $ \omega_c + j \omega_T$ has degeneracy $2(j+1)$ as one expects for a 2D harmonic oscillator. The corresponding cavity state degeneracy is illustrated in Fig.~\ref{fig:cavity}(b). Defining $n_0$ as the average photon number in the ground mode, The black lines in Fig.~\ref{fig:fractot}(a)-(b) show the numerically calculated curves of the condensate fraction $n_0/n_{\rm tot}$ and $n_{\rm tot}$ as a function of $\mu$ in the absence of loss with $T$ and $\omega_T$ fixed. In our finite system we do not have a sharp transition; 
We define the transition to BEC to be at the inflection point, where $d^2 (n_0/n_{\rm tot})/d [\log(\mu)]^2 =0$ in Fig.~\ref{fig:fractot}(a). 
Treating the total number of excited photons in the continuous limit~\cite{Mullin1997}, the phase transition according to this definition\footnote{Here we study the scenario with varying $n_{\rm tot}$ while keeping $\omega_T$ constant, which is different from the usual case to approach thermodynamic limit with a fixed number density.} occurs at the critical value $\mu = - 3 \hbar^2 \omega_T^2/ \pi^2 k_B T$, which coincides with the condition $n_0/n_{\rm tot} \approx 1/2$. The total number of photons at the transition point in this lossless limit for these parameters is $\approx 26\, 000$.
    
The position of the inflection point in $n_0/n_{\rm tot}$ [Fig.~\ref{fig:fractot}(a)] will shift in the presence of loss---due to both cavity loss and scattering into the free-space modes---whose effects become important at lower laser power. In the regime we focus on in this work, the cavity loss at the mirrors can be neglected because the effective optical depth is taken to be large enough that the cavity photons will interact with an atom before being lost at the mirrors. The effect of cavity photon loss by scattering into the free-space modes can be suppressed by increasing the cooling laser intensity.  However, higher order effects will need to be considered if one works beyond the low excitation limit $\Omega^2 \ll \abs{\bar{\Delta}_L}^2$.

For $\Omega > \Omega_c$, the loss-modified total number of photons is given by replacing
$\mu$ with $\tilde{\mu} \equiv (\hbar\omega_L-\hbar\omega_c-\delta \mu)$, which is the loss-modified displaced chemical potential, and replacing $\beta$ with $\beta_{\rm eff}[j]$ in Eq.~(\ref{ntot}). The effective temperature $\beta_{\rm eff}[j]^{-1}$ is mode-dependent, and decreases as $j=2l+\abs{m}$ increases. We study the BEC transition with scattering loss numerically as shown in the colored lines in Fig.~\ref{fig:fractot}. For scenarios with a fixed total number of photons, which is a closer analogy to atomic BEC, the condensate fraction $n_0/n_{\rm tot}$ is shown as the colored lines in Fig.~\ref{fig:fractot}(c). The modified curves resemble qualitatively the ideal grand canonical ensemble case [black lines in Fig.~\ref{fig:fractot}(c)] at large $n_{\rm tot}$ and large $\Omega/|\bar{\Delta}_L|$, but with a higher transition temperature. The increase in the transition temperature arises from the loss-induced truncation of the populations of the higher frequency modes leading to these modes no longer being in thermal equilibrium. These populations are significantly lower than would be predicted by a single temperature equal to the atom temperature (see Fig.~5 in ~\cite{Wang2018a}).  Thus, for a fixed $n_{\rm tot}$ and $T$, the mode occupation of the lower modes is significantly higher than in the untruncated case, which increases the transition temperature. Just as in a trapped atomic gas BEC, higher central density (ground mode occupation) leads to a higher transition temperature.

\begin{figure}[htbp]
\begin{center}
\includegraphics[width=\linewidth]{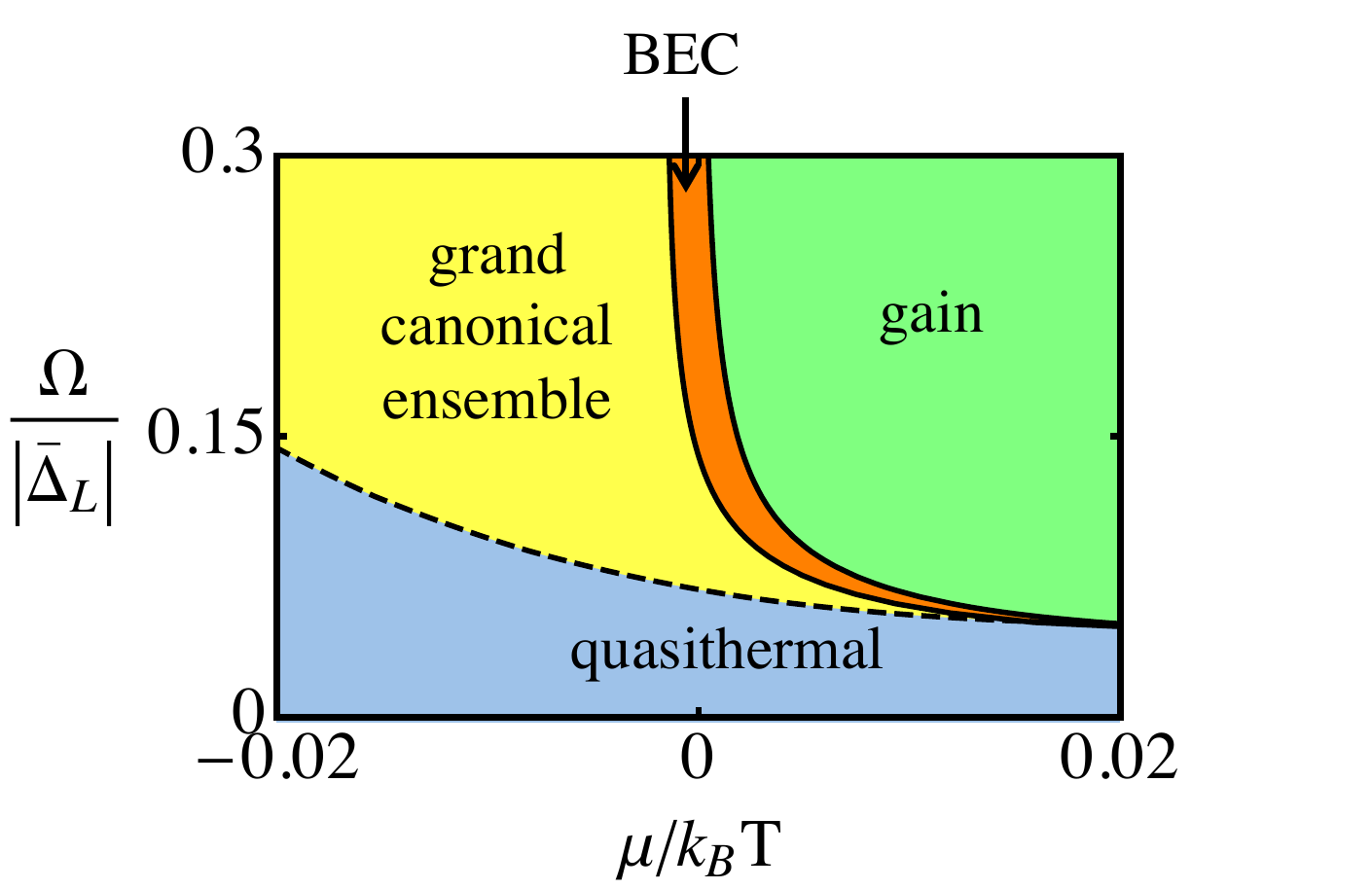}
\caption{Calculated phase diagram of cavity photons as a function of $\Omega$ and $\mu$. At high power, photon generation can exceed loss leading to gain (green region) and possibly lasing; cavity photons can be described as a grand canonical ensemble (yellow) at equilibrium, and we find the formation of a photon BEC (orange) within the GCE area and near the gain boundary.  For low power, photon loss prevents equilibration of photons to a single temperature corresponding to that of the atomic motion, and only quasithermal light (blue) is expected.  In this diagram we use the same parameters as in Fig.~\ref{fig:fractot}(a).}
\label{fig:Phase}
\end{center}
\end{figure}

The loss-modified condensate fraction and the corresponding total number of photons as a function of $\tilde{\mu}$, at fixed $T$ and $\omega_T$, are shown in colored lines in Fig.~\ref{fig:fractot}(a)-(b). The transition between solid and dotted segments marks the distinction between GCE-like and quasithermal regimes as described below. The solid segments of the colored lines are qualitatively similar to the ideal result (black) with the inflection points of Fig.~\ref{fig:fractot}(a) left-shifted, which also arises from the loss-induced truncation of the populations of the higher frequency modes. On the other hand, the dotted part of our modified result is showing drastically different features from the ideal curve: Instead of being a monotonic function of $\tilde{\mu}$, the modified $n_0/n_{\rm tot}$ reaches a minimum then eventually increases to 1 when $\tilde{\mu}$ decreases further away from zero. The total number of photons also decreases substantially in this regime.  This behavior is due to the fact that higher frequency modes have lower effective temperature because of loss; the occupation will tend toward the limit of $n_0/n_{\rm tot}=1$ for large, negative $\tilde{\mu}$ not because of a high degree of condensation but rather because only one mode survives the loss.  We again define the BEC boundary to be at the inflection points of the condensation fraction curves.  We then define an empirical condition that separates the GCE-like (solid line) region from the quasithermal (dotted line) region in Fig.~\ref{fig:fractot}(a)-(b): We define a grand canonical ensemble phase in which $-1/2 \leq \log_{10}\left(\frac{T_{\rm eff}[5]}{T_o} \right)\leq 0$, where $T_o$ is a reference temperature at the equilibrium-to-gain transition, such that there are at least $\sum_{j=0}^5 2(j+1)=42$ modes that can be effectively described by a single temperature. For larger negative $\tilde{\mu}$ or lower laser intensities, the scattering loss prevents detailed balance of the cavity photons with atomic motion, and only quasithermal light (where the photon distribution cannot be described by a single temperature even for a moderate number of modes) is expected.  For $\tilde{\mu} >0$, one expects the onset of gain for the ground mode.  The calculated phase diagram of the cavity photons is summarized in Fig.~\ref{fig:Phase} with the phase boundaries defined above.

\begin{figure}[htbp]
\begin{center}
\includegraphics[width= \linewidth]{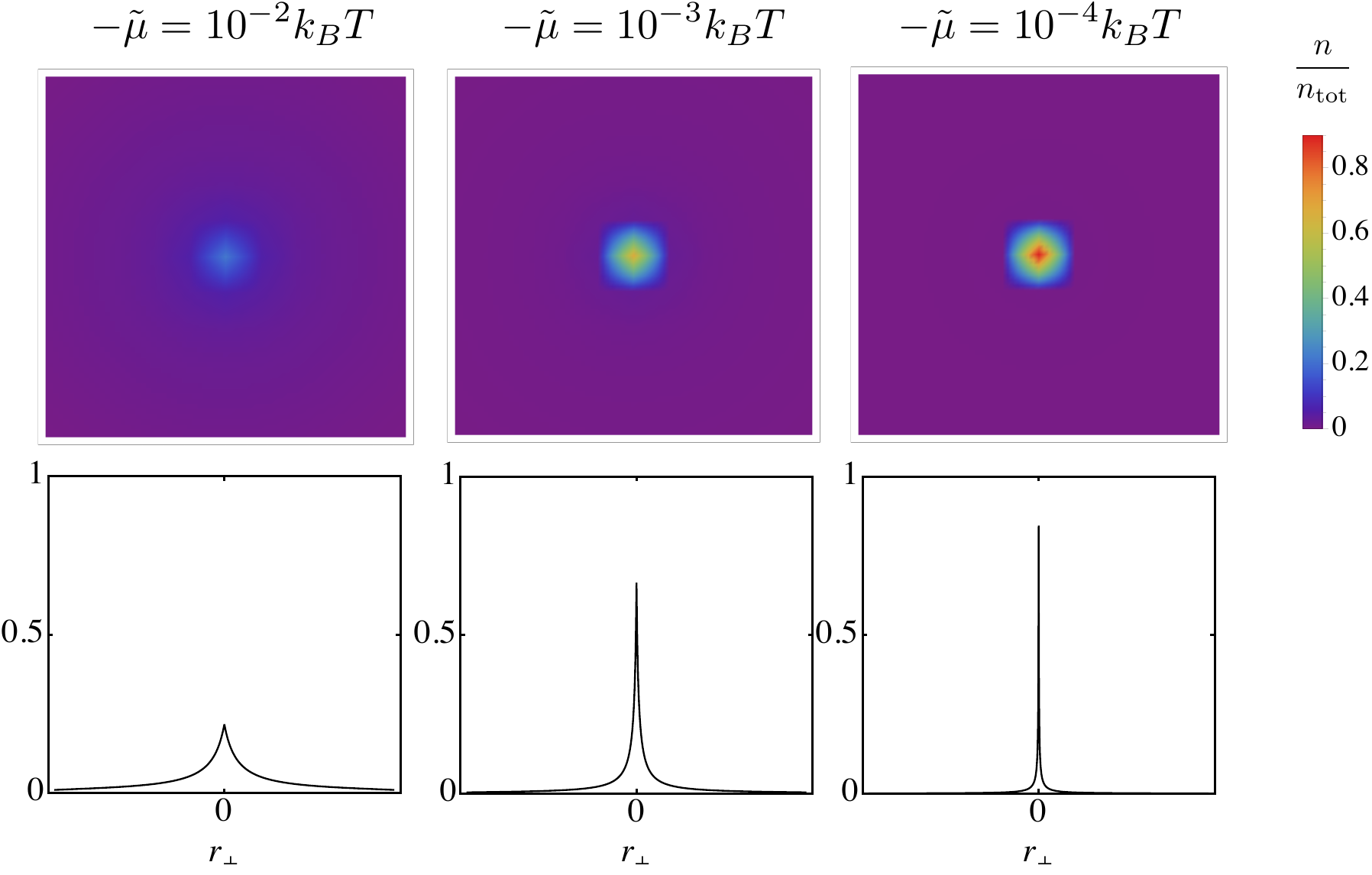}
\caption{Simulations of the far-field photonic ``time-of-flight" images and corresponding cross-sections through the center, where $r_{\perp}$ is the far-field transverse position. The parameters are the same as in Fig.~\ref{fig:fractot}(a) with $\Omega=0.3\abs{\bar{\Delta}_L}$.}
\label{fig:ToF}
\end{center}
\end{figure}

What can one observe in an experiment?  In atomic BEC experiments, a typical technique to observe a BEC transition is to use the time-of-flight method to measure the momentum distribution of atoms.
Here, the photonic version of ``time-of-flight" is the far-field distribution of light emitted from the cavity, which reflects the momentum distribution of the cavity transverse modes. Simulations of the photonic ``time-of-flight" images according to Eq.~(\ref{DetailedBalance}) are shown in Fig.~\ref{fig:ToF}.  One expects a sharp central peak when $\tilde{\mu}$ is near zero, representing condensation into the ground mode. 

\section{Summary and Outlook}
We have shown that Doppler cooling of a dilute, two-level, atomic ensemble inside an optical cavity can lead to 2D Bose-Einstein condensation of light. By studying the condensate fraction and the total photon number with values appropriate for the Yb intercombination transition, we have constructed a phase diagram as a function of laser frequency and field strength showing gain, condensate, thermal, and quasithermal regimes for cavity photons.  The simplicity as well as the high degree of control of our approach open up opportunities in exploring quantum phenomena with light.  In particular, the thermalization arguments can be directly generalized to include nonlinear interactions, and thus are relevant to applications such as Rydberg-polariton thermalization with laser-cooled Rydberg atoms, photon superfluidity, and nonequilibrium phase transitions.

\begin{acknowledgments}
We thank A. V. Gorshkov, S. M. Girvin, E. Benck, E. A. Goldschmidt, V. Vuleti\'{c}, M. Weitz, H. Carmichael, and S. Ragole for helpful discussions.
Funding is provided by NSF Physics Frontier Center, PFC @ JQI.
\end{acknowledgments}

\bibliographystyle{apsrev4-1}
\bibliography{BECmendeley}

\end{document}